\documentclass[reprint,nofootinbib,amsmath,amssymb,aps]{revtex4-1}
\usepackage{graphicx}
\usepackage{dcolumn}
\usepackage[colorlinks,linkcolor=magenta,anchorcolor=cyan,citecolor=blue]{hyperref}
\usepackage{bm}
\usepackage[multiple]{footmisc}
\newcommand{\fig}[1]{Fig.\ref{#1}}
\def\be{\begin{equation}}
\def\ee{\end{equation}}
\def\ba{\begin{eqnarray}}
\def\ea{\end{eqnarray}}

\def\nn{\nonumber}
\def\lf{\left}
\def\rt{\right}

\newcommand{\eq}[1]{(\ref{#1})}

\def\nn{\nonumber}\def\lf{\left}\def\rt{\right}  \def\t {\tau} \def\y {\psi}   \def\p {\pi} \def\a {\alpha}  \def\d {\delta} \def\f {\phi} \def\g {\gamma} \def\h {\eta} \def\j {\varphi} \def\k {\kappa} \def\l {\lambda} \def\z {\zeta} \def\x {\xi} \def\c {\chi} \def\b {\beta}  \def\m {\mu} \def\pd {\partial} \def \inf {\infty}  
\def\Q{\Theta} \def\W{\Omega}     \def\S {\Sigma} \def\D {\Delta} \def\F {\Phi}  \def\L {\Lambda}    \def\grad{\nabla}\def\.{\cdot}
\def\math {\mathcal}
\begin{document}

\title{Switchback effect of holographic complexity in multiple-horizon black holes}
\author{Jie Jiang}
\email{jiejiang@mail.bnu.edu.cn}
\author{Zhaohui Chen}
\email{chenzhaohui@mail.bnu.edu.cn}
\author{Chengcheng Liu}
\email{liucheng@mail.bnu.edu.cn}
\affiliation{Department of Physics, Beijing Normal University, Beijing 100875, China\label{addr1}}
\date{\today}

\begin{abstract}
In this paper, we use the ``complexity equals action'' (CA) conjecture to explore the switchback effect in the strongly-coupled quantum field theories with finite $N$ and finite coupling effects. In the perspective of holography, this is equivalent to evaluating the CA complexity in a Vaidya geometry equipped with a light shockwave for a higher curvature gravitational theory. Based on the Noether charge formalism of Iyer and Wald, we obtain the slope of the complexity of formation in the small and large time approximations. By circuit analogy, we show that our results concur with the switchback effect of the quantum system. These results show that the switchback effect is a general feature of the CA complexity in stationary black holes and its existence is independent of the explicit gravitational theory as well as spacetime background. From the viewpoint of AdS/CFT, this also implies that the switchback effect is a general feature of the thermofield double state in the strongly-coupled quantum field systems with finite $N$ and finite coupling effects. Moreover, we also illustrate that unlike the late-time complexity growth rate, the counterterm plays an important role in the study of the switchback effect.

\end{abstract}
\maketitle
\section{Introduction}

In recent years, quantum information perspectives have provided many useful techniques for studying the AdS/CFT correspondence. This idea has aroused more and more attention to the concept of ``quantum circuit complexity'', which is defined as the number of the elementary gates in the optimal circuit from a given state to a target state \cite{Watrous,Aaronson,Susskind01,Susskind02,Susskind03}. From the perspective of holography, two complementary conjectures for the bulk description of the complexity of boundary states have been proposed: the ``complexity equals volume'' (CV) \cite{Susskind03,D.Stanford} and the ``complexity equals action'' (CA) \cite{BrL,BrD} conjectures.  The CV conjecture states that the circuit complexity of a quantum state $|\y(t_L,t_R)\rangle$ in boundary strongly-coupled system is dual to the volume $V$ of the Einstein-Rosen bridge anchored at the time slices $t_L$ and $t_R$ on the boundary, i.e.,
\ba\begin{aligned}
C_V\lf(|\y(t_L,t_R)\rangle\rt)=\frac{V}{G\ell_\text{AdS}}\,.
\end{aligned}\ea
On the other hand, the CA conjecture states that the complexity of boundary state is given by evaluating the full on-shell action of the bulk gravitational theory on the Wheeler-DeWitt (WDW) patch, which is the causal development of a spacelike bulk surface (Cauchy surface) connected the boundary timeslices $t_L$ and $t_R$, i.e.,
\ba\label{CA}
C_A\lf(|\y(t_L,t_R)\rangle\rt)\equiv\frac{I_\text{WDW}}{\p\hbar}\,.
\ea
{ These conjectures have attracted researchers to study the complexity of the strongly-coupled quantum system from the perspective of the holographic principle \cite{Susskind1,Susskind2,Susskind3,Susskind4,Susskind5,Susskind6,Zhao1,Zhao2,Zhao3,Myers1,Myers2,Myers3,Myers4,Myers5,Myers6,Myers7,Chapman1,
Chapman2,Fan1,Fan2,Fan3,Fan4,Jiang1,Jiang2,Jiang3,Jiang4,Jiang5,Jiang6,Jiang7,Jiang8,An1,An2,An3,Yang1,Yang2,Yang3,Yang4,Yang5,Yang6,1,2,3,4,5,6,7,8,
9,10,11,12,13,14,15,16,17,Moosa,Nally:2019rnw,Auzzi:2019mah,Tanhayi:2018gcj,s1,s2,s3,s4}.}

{
The concepts of the local and global quantum quenches have been widely used to study the holographic complexity. The local quench is the process when the system evolves after a local perturbation. It was argued that the holographic dual of this process is given by the black hole geometry perturbed by the particle falling on the horizon \cite{local1,local2,local3,local4,local5}. Based on this setup, Lloyd's bound \cite{Lloyd} of the complexity in the boundary system has been tested under the local quench by using the holographic conjectures\cite{LQ1, LQ2, LQ3}. If the perturbation is global, the process is called a global quench. The holographic dual of this process is given by the Vaidya geometry, which is equipped with a thin shell of null fluid collapse (shockwave) \cite{G1, G2, G3, G4, G5, G6, G7, G8}. Based on this duality, the time-dependence of the complexity in the boundary quantum field system has been studied by using different holographic complexities \cite{Chapman1, Chapman2, Moosa, Fan2, Jiang2, Yang3, Auzzi:2019mah, Tanhayi:2018gcj}.
}

It has been generally argued in \cite{D.Stanford, Susskind5} that the quantum complexity in chaos system should exhibit the switchback effect when this system is perturbed by a quantum quench, which can be described by a precursor $\math{O}(t)=U^\dag(t)\math{O} U (t)$ of a simple perturbation operator $\math{O}$. Since $\math{O}$ is a very simple operator, at a very early time, it can be regarded as a unit operator and then the complexity does not grow. However, for a chaotic system, at a very large time compared to the scrambling time $t_\text{scr}^\star$, the operator $\math{O}$ will disrupt the time-reversed evolution and the complexity will become twice the complexity of the evolution operator $U(t)$ \cite{D.Stanford}. This property of the quantum complexity under the quantum quench is known as the switchback effect \cite{D.Stanford}. This feature plays an important role to examine the definitions the complexity.

{  Although lots of researchers focused on the calculation of the circuit complexity in quantum field theory \cite{jiang1,jiang2,myers1,myers2,myers3,myers4,yang1,yang2,yang3,yang4,f1,f2,f3,f4}, there is still a lack of a valid method to evaluate the circuit complexity in the strongly-coupled system. Therefore, some researchers used the holographic complexity in the Vaidya black holes for Einstein gravity to study the switchback effect in the boundary strongly-coupled system \cite{Susskind5, D.Stanford, Chapman2, Jiang2}. However, in the context of AdS/CFT, the Einstein gravity in bulk is dual to the strongly-coupled quantum field with infinite $N$ or infinite coupling effects. A natural question for us is to ask whether the switchback effect is a general feature of the strongly-coupled quantum system and independent on the explicit of the quantum state as well as the quantum theory. As we all know, the boundary quantum field theory with finite $N$ and finite coupling effects are corresponding to a gravitational theory with higher curvature corrections. Therefore, in this paper, we would like to use the CA conjecture to show whether the switchback effect also exists in the strongly-coupled quantum system with finite $N$ and finite coupling effect under a global quantum quench.}

Recently, some authors found that once the higher curvature corrections are taken into account, the neutral (single-horizon) black holes will have a divergent complexity growth rate since the WDW patch will go arbitrarily near the singularity \cite{Nally:2019rnw,Jiang8}. However, this does not happen for multiple-horizon black holes due to the different causal structure including at least two horizons. Moreover, most of the previous researches imply that the CA complexity for the neutral black hole can be obtained by taking the limit of its corresponding multiple-horizon counterpart \cite{Yang6,s1,s2,s3,s4}. Therefore, in order to reflect some universal features of the CA complexity and avoid the divergent result of the neutral case, in this paper, we would like to focus on the black holes which have at least two Killing horizons. Generally, these black holes capture some extra conserved charges, such as the angular momentum and electric charge. From the viewpoint of the holography, they are dual to the boundary quantum state which also contains some extra conserved charges. Therefore, our investigations can also reflect the influence of the extra conserved charges on the complexity in the boundary quantum system.

The above statements show that the main task of this paper is to evaluate the CA complexity in the Vaidya geometry equipped with a light shockwave. By analyzing this geometry in Sec.\ref{Sec1}, we can see that at the large time limit, the dynamical points will approach the Killing horizons. Then, the actions that we need to evaluate are in the regions which are connected to the Killing horizons and can be generated by the corresponding Killing vectors. This property allows us to express this action as some boundary integrals based on Iyer-Wald formalism \cite{IW}. Therefore, in the following, we would like to utilize the Iyer-Wald formula to derive some general expressions of the CA complexity at large times.

The remainder of our paper is organized as follows: in Sec. \ref{Sec1}, we first study the geometry of the stationary black hole with a thin shell of null fluid collapse. In Sec. \ref{Sec2}, we briefly review the Iyer-Wald formalism for a invariant theory. In Sec. \ref{Sec3}, we investigate the switchback effect of the CA complexity in a multiple-horizon black hole for a general higher curvature gravitational theory coupled with arbitrary matter fields. In Sec. \ref{Sec4}, we compare our holographic results with the circuit behaviors.

\begin{figure}
\centering
\includegraphics[width=0.45\textwidth]{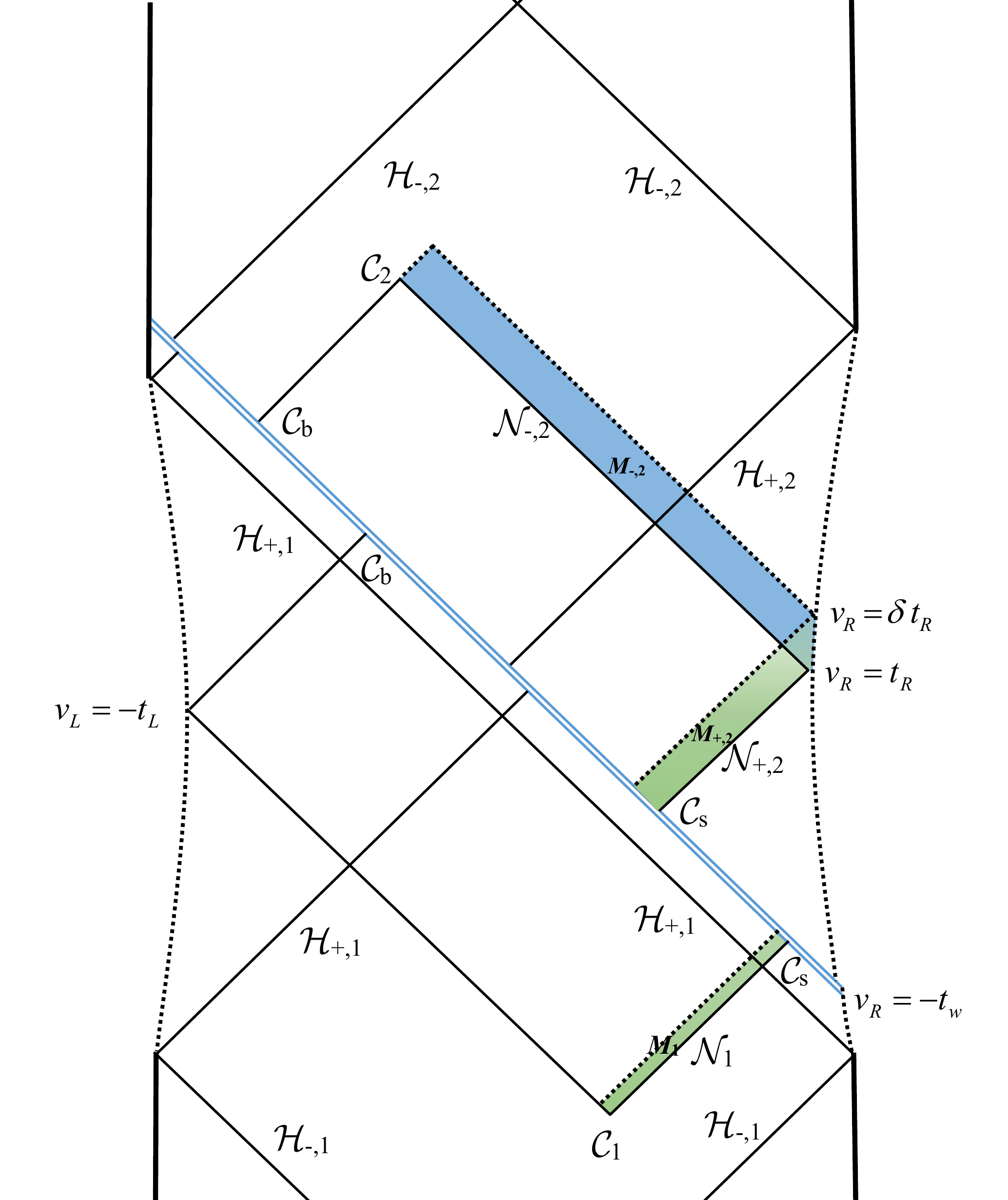}
\caption{Wheeler-DeWitt patch of a multiple Killing horizon black hole which is equipped with a thin shell of null fluid collapse, where the dashed lines denote the cut-off surface at asymptotic infinity, satisfying the asymptotic symmetries. }\label{WDW}
\end{figure}

\section{Geometry with a large time light shockwave}\label{Sec1}

As mentioned above, the stationary multiple horizons of the black holes are caused by the extra conserved charges of the spacetime. In the context of AdS/CFT, these black holes are dual to the thermofield double (TFD) states
\ba\begin{aligned}
|\text{TFD}\rangle=Z^{-1/2}\sum_{\a}e^{-\b(E_\a+\m Q_\a)/2}|E_\a,Q_\a\rangle_\text{L} |E_\a,-Q_\a\rangle_\text{R}
\end{aligned}\nn\\\ea
on the boundary strongly-coupled quantum field theory \cite{TFD}, where we have denoted the subscripts L and R to the left and right boundaries of the multiple-horizon black hole geometry individually. Here $E_\a$ and $Q_\a$ are corresponding to the eigenvalues of the energy and extra conserved charges separately. The time evolution of the TFD state is obtained by
\ba\begin{aligned}
|\text{TFD}(t_L,t_R)\rangle=U_L(t_L)U_R(t_R)|\text{TFD}\rangle\,.
\end{aligned}\ea
where
\ba\begin{aligned}
U_L(t_L)=e^{-i (H_L+\m Q_L)t_L}\,, U_R(t_R)=e^{-i (H_R-\m Q_R)t_R}
\end{aligned}\ea
are the time evolution operators corresponding to the left and right quantum system. We can see that this state is invariant under the shift transformation
\ba\begin{aligned}
t_L\to t_L+\D t\,,\ \ \ \ \ t_R\to t_R-\D t\,.
\end{aligned}\ea

In this paper, we would like to investigate the switchback effect of the complexity. Following the setup in Refs. \cite{D.Stanford,Susskind5,Chapman2}, we consider the perturbation of the TFD state,
\ba\begin{aligned}
|\text{TFD}\rangle_\text{pert}&=\math{O}_R(-t_w)|\text{TFD}\rangle\\
&=U_R(t_w)\math{O}_RU_R^\dag(t_w)|\text{TFD}\rangle
\end{aligned}\ea
where the precusor $\math{O}_R(-t_w)=U_R(t_w)\math{O}_RU_R^\dag(t_w)$ is the perturbation operator inserted into the right sight quantum system at time $-t_w$, and $\math{O}_R$ is a localized simple operator. In the chaos quantum system, $U_R(t_w)$ and $U_R^\dag (t_w)$ will approximately cancel until times of the order of the scrambling time $t_\text{scr}^\star$ \cite{D.Stanford} and the state would keep unchanged, i.e., the complexity growth rate is vanishing. For $t_w\gg t_\text{scr}^\star$, the complexity growth rate of $\math{O}(t_w)$ is just twice the rate of the evolution operator $U_R(t_w)$. This nontrivial feature is connected to the switchback effect \cite{D.Stanford,Susskind5}. Evolving the perturbed state in the right and left time gives
\ba\begin{aligned}
|\text{TFD}(t_L,t_R)&\rangle_\text{pert}=U_L(t_L)U_R(t_R)|\text{TFD}\rangle_\text{pert}\\
&=U_R(t_R+t_w)\math{O}_RU_R(t_L-t_w)|\text{TFD}\rangle\,.
\end{aligned}\ea

In the holographic context, the dual geometry in the bulk to above perturbed system is AdS-Vaidya spacetime with multiple horizons source by a thin shell of null fluid collapses. It can be described by the metric ansatz
\ba
ds^2=ds^2_2\math{H}(v+t_w)+ds^2_1\lf[1+\math{H}(v+t_w)\rt],
\ea
where $\math{H}(v)$ is the Heaviside step function, and
\ba\begin{aligned}\label{ds2}
&ds^2_{i}=\a_i(r,y)\lf[-f_i(r)dv^2+2dv dr\rt]\\
&+\g_{AB}^{(i)}(r,y)\lf[dy^A-\W^{(A)}_i(r,y)dv+\frac{\W^{(A)}_i(r,y)}{f_i(r)}dr\rt]\\
&\times \lf[dy^{(B)}-\W^{(B)}_{i}(r,y)dv+\frac{\W^{(B)}_{i}(r,y)}{f_i(r)}dr\rt]\,,
\end{aligned}\ea
{
where $i=1,2$ describes a multiple-horizon stationary black hole, the indexes $A, B$ denote the coordinates of the codimension-two surface, $\W^{(A)}$ is some component of the metric which becomes the angular velocity of the black holes when it evaluates on the Killing horizon.
The metric in \eq{ds2} is a generalization from most stationary axisymmetric black holes in general relativity or other theories of gravity, such as:
Kerr-AdS(dS) black holes, Myers-Perry black holes \cite{Myers:1986un}, rotating Bardeen black holes \cite{Bambi:2013ufa}, rotating Hyward black holes \cite{Bambi:2013ufa}, rotating charged cylindrical black holes \cite{Lemos:1995cm}, Kerr-MOG black holes \cite{Moffat:2014aja}, Kerr-Sen black holes\cite{Sen:1992ua}, Kerr-Newman-Taub-NUT-AdS black holes \cite{Taub1}, Gauss-Bonnet black holes \cite{GB}, rotating black holes in a Randall-Sundrum brane \cite{RS}, and charged accelerating AdS black holes \cite{caads}. }

By virtue of the second law of thermodynamics for black holes, we will set $r_{+,2}>r_{+,1}$ with horizon radius $r_{\pm,i}$ determined by $f_i(r_{\pm,i})=0$. This line element describes an infinitely thin shell collapse which generates a shape transition from a black hole with the metric $ds_1^2$ to another one with $ds^2_2$.  For the convenience of later calculations, we would like to introduce the tortoise coordinates as
\ba\begin{aligned}
v_R&<-t_w:\ \ \ \ \ r_1^*(r)=\int \frac{dr}{f_1(r)}\,,\\
v_R&>-t_w:\ \ \ \ \ r_2^*(r)=\int \frac{dr}{f_2(r)}\,.
\end{aligned}\ea
When the position considered is near the horizon, i.e., $r\simeq r_{\pm,i}$, we have
\ba\label{nearrstar}
r_i^*(r)\simeq \frac{1}{2\k_{\pm,i}}\ln\left|\frac{r-r_{\pm,i}}{r+r_{\pm,i}}\right|\,,
\ea
where
\ba\begin{aligned}
\k_{\pm,i}=\frac{|f'_i(r_\pm)|}{2}
\end{aligned}\ea
is the surface gravity corresponding to the Killing horizons. Using these coordinates, one can also define an ``outgoing" null coordinate $u$ and auxiliary time coordinate $t$ as
\ba
u_{i}\equiv v-2r^*_i(r),\ \ \ \ \ \ \ t_i\equiv v-r^*_i(r)\,.
\ea

According to the CA conjecture \eq{CA}, computing the quantum complexity of the boundary state is equivalent to evaluating the full action within the WDW patch. As is shown in \fig{WDW}, the geometry of the WDW patch is characterized by some dynamical points: $r_1$ and $r_2$, the points where the past/future null boundaries of the WDW patch meet inside the horizon; $r_s$ and $r_b$, the positions where the right past and left future boundaries meet the shockwave. Moreover, in order to regulate the divergence caused by the asymptotic infinity, a cut-off surface $r=r_\L$ is also introduced.

Performing the tortoise coordinates, one can find that these dynamical positions $r_s,r_b,r_1$ and $r_2$ yield
\ba\begin{aligned}\label{rrrr}
t_w+2r_2^*(r_s)&=-t_R\,,\\
t_w+2r_1^*(r_b)&=t_L\,,\\
t_w+2r_1^*(r_s)&=t_L+2r_1^*(r_1)\,,\\
t_w+2r_2^*(r_b)&=-t_R+2r_2^*(r_2)\,.\\
\end{aligned}\ea
For the cases with light shockwave, there exists a scrambling time
\ba
t_\text{scr}^* =\frac{1}{2\p T_{+,1}}\ln \frac{2}{\d}\,,
\ea
which divides the evolution into two asymptotic regions. Here we have denoted
\ba\begin{aligned}
T_{\pm,i}=\frac{1}{4\p}f_i'(r_{\pm,i})\,,\ \ \d=\frac{r_{+,2}}{r_{+,1}}-1\,.
\end{aligned}\ea

For the case with a light shockwave, the scrambling time becomes very large. According to the expressions \eq{rrrr}, we can see that the dynamical point $r_s$ approaches the horizon $\math{H}_{2,+}$. Then, when $t_w-t_L > t_\text{scr}^*$ and $t_w+t_R > t_\text{scr}^*$, we have
\ba\begin{aligned}
2r_1^*(r_1)&=t_w-t_L-t_\text{scr}^*\,,\\
2r_2^*(r_2)&=t_w+t_R-t_\text{scr}^*\,.
\end{aligned}\ea
These expressions imply that the scrambling time $t_\text{scr}^*$ is a transition position for $r_i$ between $r_{+,i}$ and $r_{-,i}$. Then, in the limit of large $t_w$, we have $r_s\to r_{+,2}$, $r_b\to r_{+,1}$ and $r_i\to r_{-,i}$.

Finally, we consider the behaviors of the null segment which crosses the shockwave.  According to the line element \eq{ds2}, it is easy to check that
\ba\begin{aligned}
l_a=l^a_2\math{H}(r-r_s)+\L_s l_1^a\lf[1-\math{H}(r-r_s)\rt]\,,\\
\tilde{l}_a=\tilde{l}^a_1\math{H}(r-r_b)+\L_b \tilde{l}_2^a\lf[1-\math{H}(r-r_b)\rt]\,,\\
\end{aligned}\ea
with
\ba\begin{aligned}
l_i^a=-(dv)_a+\frac{2}{f_i(r)}(dr)_a\,,\ \ \ \ \L_s=\frac{f_1(r_s)}{f_2(r_s)}\,,\\
\tilde{l}_i^a=-(dv)_a+\frac{2}{f_i(r)}(dr)_a\,,\ \ \ \ \L_b=\frac{f_2(r_b)}{f_1(r_b)}\,,
\end{aligned}\ea
are the affine null generator of the past right and future left null boundaries, individually. We can find that $\math{L}_\z l^a_i=0$ if $\z^a$ is a Killing vector field, such as the stationary Killing vector $t^a=(\pd/\pd v)^a$ or the axial Killing vector $\j^a_{(\m)}$. Then, using Eq. \eq{rrrr}, one can further obtain
\ba\begin{aligned}\label{tj}
&\frac{d\ln \L_s}{d t_R}\simeq-2\p T_{+,2}\,,\ \ \ \ \frac{d u_{2,s}}{d t_R}=\L_s^{-1}\simeq0\,,\\
&\frac{d\ln \L_b}{d t_L}\simeq2\p T_{+,1}\,,\ \ \ \ \ \ \ \frac{d u_{1,b}}{d t_L}=-\L_b^{-1}\simeq0\,,
\end{aligned}\ea
at the large time limit. These results imply that the past right null segment before shockwave as well as the future right null segment after shockwave keep almost unchanged when we vary the left or right boundary times.

\section{Iyer-Wald formalism}\label{Sec2}
According to the discussion in the last section, we can see that all of the dynamical points of WDW patch are located on the horizons $\math{H}_{\pm,i}$ in the limit of large time $t_w$. The integral region for calculating the change of the bulk action can be generated by the diffeomorphism related to the Killing vector field of the Killing horizon. Using the Stokes' theorem, we can express these action as some boundary integrals related to the Killing vector fields. On the other hand, Iyer and Wald \cite{IW} perform the differential form to obtain the relationship between the conserved charges related to some vector fields and the action integrals. Therefore, it might be possible for us to derive some general expressions of the CA complexity at the large time limit based on the Iyer-Wald formalism. Next, we would like to give a brief review of the Iyer-Wald formalism for a general diffeomorphism invariant theory, which is described by a Lagrangian $\bm{L}=\math{L}\bm{\epsilon}$ where the dynamical fields consist of a Lorentz signature metric $g_{ab}$ and other fields $\y$. Following the notation in \cite{IW}, we use boldface letters to denote differential forms and collectively refer to $(g_{ab},\y)$ as $\f$. Generally, the action can be divided into the gravity part and matter part, i.e., $\bm{L}=\bm{L}_\text{grav}+\bm{L}_\text{mt}$. The variation of the gravitational part with respect to $g_{ab}$ is given by
\ba
\d \bm{L}_\text{grav}=\bm{E}_{g}^{ab}(\f)\d g_{ab}+d \bm{\Q}(\f,\d g)\,,
\ea
where $\bm{E}_g^{ab}(\f)$ is locally constructed out of $\f$ and its derivatives and $\bm{\Q}$ is locally constructed out of $\f,\, \d g_{ab}$ and their derivatives. The equation of motion can be read off as
\ba
\bm{E}_g^{ab}(\f)=\frac{1}{2}T^{ab}\bm{\epsilon}\,,
\ea
where
\ba\label{Tab}
T^{ab}=-\frac{2}{\sqrt{-g}}\frac{\d \sqrt{-g}\math{L}_\text{mt}}{\d g_{ab}}=-g^{ab}\math{L}_\text{mt}-2\frac{\d \math{L}_\text{mt}}{\d g_{ab}}
\ea
is the stress-energy tensor of the matter fields. Let $\z^a$ be the infinitesimal generator of a diffeomorphism. Exploiting the Bianchi identity $\grad_{a}T^{ab}=0$, one can obtain the identically conserved current for a generic background metric $g_{ab}$ as
\ba\label{J1}\begin{aligned}
\bm{J}[\z]&=\bm{\Q}(\f,\z)-\z\.\bm{L}_\text{grav}
+s_{\z}\.\bm{\epsilon}\,,
\end{aligned}\ea
 where $s^a_{\z}\equiv T^{ab}\z_b$ and  $\bm{\Q}(\f,\z)=\bm{\Q}(\f,\math{L}_{\z}g_{ab})$. Since $\bm{J}$ is closed, there exists a Noether charge $(n-2)$-form $\bm{K}[\z]$ such that $\bm{J}[\z]=d\bm{K}[\z]$. With similar arguments in \cite{IW}, this $(n-2)$-form can always be expressed as
\ba\label{K}
\bm{K}=\bm{W}_c\z^c+\bm{X}^{cd}\grad_{[c}\z_{d]}\,,
\ea
where
\ba\label{Xcd}
\lf(\bm{X}^{cd}\rt)_{c_3\cdots c_n}=-E_R^{abcd}\bm{\epsilon}_{abc_3\cdots c_n}\,,
\ea
is the Wald entropy density in which
\ba
E_R^{abcd}=\frac{\d \math{L}_\text{grav}}{\d R_{abcd}}\,.
\ea

Particularly, when $\z^a$ is taken to a rotational Killing vector $\j^a$ in an axisymmetric spacetime, by using this $(n-2)$-form, we can construct a conserved charge
\ba\label{JK}
J[\j]=-\int_{\math{C}_\inf}\bm{K}[\j]\,,
\ea
where $\math{C}_\inf$ denotes a $(n-2)$-dimensional surface at the asymptotic infinity.  It can be interpreted as the angular momentum of the black hole in an arbitrary asymptotic space\cite{WSS}. For a general higher curvature gravitational theory, it is given by
\ba\label{Kj}
J[\j]=-\int_{\math{C}_\inf}\lf(\bm{X}^{cd}\grad_{[c}\j_{d]}-2\j_{b}\grad_a\bm{X}^{ab}\rt)\,.
\ea

Moreover, if we set $\z^a$ to a Killing vector, by substituting (\ref{Tab}) into (\ref{J1}), one can obtain
\ba\label{Lz}
\z\.\bm{L}=d\lf(\bm{\L}[\z]-\bm{K}[\z]\rt)\,,
\ea
where $\bm{\L}[\z]$ is a $(n-2)$-form and constructed by
\ba\label{cz}
\c_{\z}\.\bm{\epsilon}=d\bm{\L}[\z]\,,\ \text{with}\ \ \c^a_\z=-2\frac{\d \math{L}_\text{mt}}{\d g_{ab}}\z_b\,.
\ea

\section{The slope of complexity of formation}\label{Sec3}
In this section, we start to evaluate the derivative of the complexity of formation with respect to $t_w$ (the slope of the complexity of formation). Here the complexity of formation is defined as the extra complexity required to prepare the two copies of the quantum field theory in the TFD state compared to simply preparing each of the copies in the vacuum state, i.e.,
\ba
\D C=C\lf(|\text{TFD}\rangle\rt)-C\lf(|0\rangle_L\otimes|0\rangle_R\rt)
\ea

In the context of AdS/CFT, it is dual to the difference between the holographic complexity for a black hole and that for two copies of the vacuum geometry at $t_R=t_L=0$. Therefore, in the following, it is sufficient to restrict our attention to the case $t_L=t_R=0$. By considering the shift symmetry to the antisymmetric time evolution of the complexity, i.e.,
\ba
t_R\to t_R-\d t\,,\ \ t_L\to t_L+\d t\,,\ \ t_w\to t_w+\d t\,,
\ea
we can further obtain
\ba\label{DDStw}
\frac{d\D C}{d t_w}=\left[\frac{dC_A}{d t_R}-\frac{dC_A}{d t_L}\right]_{t_L=t_R=0}\,,
\ea
where we have used the fact that the complexity in vacuum geometry is time-independent. Then, using CA conjecture \eq{CA}, obtaining the slope of the complexity of formation amounts to finding the change of the full action $I$ within the WDW patch, i.e.,
\ba\begin{aligned}
\d I_L&=I(t_L+\d t_L,t_R)-I(t_L,t_R)\,,\\
\d I_R&=I(t_L,t_R+\d t_R)-I(t_L,t_R)\,.
\end{aligned}\ea
For a general higher curvature gravity, the full action can be expressed as \cite{Jiang3}
\ba\label{fa}\begin{aligned}
I=&\int_{M} \bm{L}+\int_{\math{C}}\bm{s}\h
+\int_{\math{N}} d\l\bm{s}\k+\int_{\math{N}} d\l \pd_\l\bm{s}\ln\lf(l_\text{ct}\Q\rt),
\end{aligned}\ea
where $\bm{s}=\bm{X}^{cd}\bm{\epsilon}_{cd}$ is the Wald entropy density, $\l$ is the parameter of the null generator $k^a$ on the null segment, $\k$ measures the failure of $\l$ to be an affine parameter which is derived from $k^a\grad_ak^b=\k k^b$, $\Q=\nabla_ak^a$ is the expansion scalar, and $l_\text{ct}$ is an arbitrary length scale.

As mentioned in the last section, there are two asymptotic regions: $t_w \ll t^*_\text{scr}$ and $t_w \gg t^*_\text{scr}$. In the first region with $t_w\ll t^*_\text{scr}$, we can simply approximate $ds_1^2\simeq ds_2^2$ at the limit of light shockwave, i.e., the complexity of formation is same as the unperturbed geometry. Then, by utilizing the shift symmetry, the slope of complexity of formation vanishes, i.e.,
\ba\label{ec}
\lf.\frac{d\D C}{dt_w}\rt|_{t_w\ll t^*_\text{scr}}=0\,.
\ea

Then, we consider the second region with $t_w \gg t^*_\text{scr}$. Under this limit, the joints $\math{C}_i$ and $\math{C}_s$ approach the inner horizon $\math{H}_{-,i}$ and the outer horizon $\math{H}_{+,2}$ respectively, and the left future and right past boundary of the WDW patch become the segment of the inner horizon.

To calculate the action changes at the large times,  we first focus on $\d I_{R}$ where we fix the left boundary time $t_L$ and vary $t_R$ in the right boundary.\\
\\
\noindent
\textbf{Bulk contributions}

For the bulk contributions, in the limit of the large $t_w$, according to \eq{tj}, we can see that the null segment $\math{N}_1$ keeps almost unchanged when we vary $t_R$. This implies that all of the bulk contributions only come from the bulk regions $M_{\pm,2}$, which can be generated by the Killing vector
\ba
\x_{\pm,2}^a=t^a+\W^{(A)}_\pm \j^a_{(A)}
\ea
of the Killing horizon $\math{H}_{\pm,2}$ through the null boundary $\math{N}_{\pm,2}$ of the WDW patch. For simplification, we suppress the index $\{\pm,2\}$ in the following calculation. Then, the bulk contribution from the bulk region $M_{\pm,2}$ can be written as
\ba
I_{M}=\int_{M}\bm{L}=\d t_R\int_{\math{N}}\x\cdot\bm{L}\,.
\ea
According to (\ref{Lz}), one can obtain
\ba\label{xL}\begin{aligned}
\int_{\math{N}}\x\.\bm{L}&=-\int_{\math{N}}d \bm{K}[\x]+\int_{\math{N}}d\bm{\L}[\x]\\
&=-\int_{\inf} \bm{K}[\x]+\int_{\math{C}} \bm{K}[\x]+\int_{\math{N}}d\bm{\L}[\x]\\
&=\W^{(A)}J_{(A)}-\int_{\math{C}_\inf} \bm{K}[t]+\int_{\math{C}} \bm{K}[\x]+\int_{\math{N}}d\bm{\L}[\x]\,,
\end{aligned}\ea
where the $(n-2)$-surface $\math{C}$ is the boundary of null segment $\math{N}$ near the horizon.

Since the Killing horizon contains a bifurcate surface, the first term in \eq{K} vanishes. Then, one can find
\ba
\bm{K}=\bm{X}^{cd}\grad_{[c}\x_{d]}=\k\bm{s}
\ea
on the horizon $\math{H}$, where $\bm{\epsilon}_{ab}$ is the binormal of surface $\S$, and $\k$ is the surface gravity of the horizon which satisfies $\x^a\grad_a\x^b=\k\x^b$. With these in mind, Eq. \eq{xL} becomes
\ba
\int_{\math{N}}\x\.\bm{L}=\W^{(\m)}J_{(\m)}+T S-\int_{\inf} \bm{K}[t]
+\int_{\math{N}}d\bm{\L}[\x]\nn\\
\ea
with the entropy $S_{\pm,i}=2\p\int_{\math{C}_{\pm,i}}\bm{s}$ and the temperature $T=\k/2\p$ of the corresponding horizon. Considering these relations, we have
\ba
\d I_{M_R}=\d t_R\left[\W^{(A)}J_{(A)}+T S+\L_{\inf}[\x]-\L_{\math{C}}[\x]\right]^{-,2}_{+,2}\,,
\ea
where we denoted $\d I_{M_R}= I_{M_{-,2}}-I_{M_{+,2}}$ and
\ba
\L_{\math{S}}[\z]=\int_{\math{S}}\bm{\L}[\z]
\ea
with any $(n-2)$-surface $\math{S}$ and $\math{C}_{\pm,i}$ is a codimension-$2$  section on the horizon $\math{H}_{\pm,i}$. Here the index $\{\pm,2\}$ presents the quantities evaluated at the ``outer'' or first ``inner'' horizons $\math{H}_{\pm,2}$.\\
\\
\noindent
\textbf{Surface contributions}

Next, we consider the surface contributions. Without loss of generality, we shall adopt the affine parameter for the null generator of the null surface. As a consequence, the surface term vanishes on all null boundaries. Meanwhile, by virtue of $\math{L}_{\x_{\pm,2}}l^a_2=0$, the time derivative of the counterterm contributed by $\math{N}_{\pm,2}$ vanishes. By considering that the entropy is a constant on the Killing horizon, i.e., $\math{L}_{\x}\bm{s}=0$, the counterterm contributed by the null segment on the horizon also vanishes. Therefore, the nonvanishing contribution only comes from the null segment $\math{N}_1$. And it can be written as
\ba\begin{aligned}
&I_\text{ct}(\math{N}_1)=\int_{\math{N}_1}d\t
(\pd_\t\bm{s})\ln\lf[\grad_a l^a\rt]\\
&=\int_{\math{N}_1}d\t_1(\pd_{\t_1}\bm{s})\ln\lf[\L_s\Q_1\rt]\\
&=\int_{\math{N}_1}d\t_1(\pd_{\t_1}\bm{s})\ln\Q_1-\frac{S_{-,1}}{2\p} \ln\Q_1+\frac{S_{s,1}}{2\p} \ln\Q_1\,,
\end{aligned}\ea
where we denote $\Q_1=\grad_a l^a_1$, $l^a=(\pd/\pd \t)^a$ and $l^a_1=(\pd/\pd \t_1)^a$
. Then, when we vary the right boundary, we have
\ba\begin{aligned}
\d I_\text{ct}(\math{N}_1)&=\int_{\d\math{N}_1}d\t_1(\pd_{\t_1}\bm{s})\ln\Q_1+ \d t_R T_{+,2}(S_{-,1}-S_{s,1})\\
&=\d t_R T_{+,2} (S_{-,1}-S_{s,1})\nn
\end{aligned}\\
\ea
where we used the light shockwave limit as well as the feature that $\math{N}_1$ keep unchanged at the large $t_w$.

\noindent
\textbf{Corner contributions}

Ultimately, we consider the contributions from the joints $\math{C}_{1,2}$. The affinely null generator on the horizon can be constructed as $k^a=e^{-\k\l}\x^a$ with $\x^a=(\pd/\pd \l)^a$. Then, the transformation parameter can be shown as\cite{Jiang3}
\ba
\h(\l)=\ln\lf(-\frac{1}{2}k\.l\rt)=-\k\l+\ln\lf(-\frac{1}{2}\x\.l\rt)\,.
\ea
First, we consider the corner contribution from $\math{C}_2$. Here, the change of this corner can be realized by the transformation of the killing vector $\x^a$, Then, we have
\ba
\frac{d I_{\math{C}_2}}{d t}=\frac{d I_{\math{C}_2}}{d \l_2}=-T_{-,2} S_{-,2}\,,
\ea
where we have used $\math{L}_{\x_{2,-}} l_2=0$\,. For the corner contribution from $\math{C}_1$, we have
\ba\begin{aligned}
&I_{\math{C}_1}=\int_{\math{C}_1}\bm{s}\ln\lf(-\frac{\L_s}{2}l_1\cdot k_{-,1}\rt)\\
&= \frac{S_{-,1}}{2\p}\ln\L_s+\int_{\math{C}_1}\bm{s}\ln\lf(-\frac{l_1\cdot k_{-,1}}{2}\rt)\,.
\end{aligned}\ea
Then, the change of this term becomes
\ba
\d I_{\math{C}_1}=-\d t_R T_{+,2}S_{-,1}\,,
\ea
where we considered that $\math{C}_1$ also keeps unchanged at the large time. Finally, according to
\ba
\d I=\d I_{M_R}-\d I_\text{ct}(\math{N}_1)+\d I_{\math{C}_2}-\d I_{\math{C}_1}
\ea
and the CA conjecture \eq{CA}, we can further obtain
\ba\label{dIdt}\begin{aligned}
\frac{d C_A}{dt_R}&=\math{R}=\frac{1}{\p\hbar}\left[\W^{(A)}J_{(A)}+\L_{\inf}[\x]-\L_{\math{C}}[\x]\right]^{-}_{+}\,,
\end{aligned}\ea
in the light shockwave case. Here the quantities without the index $i$ present the counterparts without the shockwave. This is actually the late-time CA complexity growth rate in the multiple-horizon black hole for a higher curvature gravity \cite{Jiang1}. When the matter fields are composed of a $U(1)$ gauge field and its corresponding complex scalar field, it will become\cite{Jiang1}
\ba
\math{R}=\frac{1}{\p\hbar}\left[\W^{(A)}J_{(A)}+\F_{H}Q_\math{C}\right]^{-}_{+}
\ea
where $\F_{\math{H}_{\pm}}$ and $Q_{\math{C}_{\pm}}$ are the chemical potential and the charge of horizon $\math{H}_\pm$, separately.

With similar calculation, we can also obtain $\d I_L=-\d I_R$. Using Eq. \eq{DDStw}, we can further obtain
\ba\label{lc}
\lf.\frac{d\D C}{d t_w}\right|_{t_w\gg t_\text{scr}^*}=2 \math{R}\,,
\ea
which is essentially twice the late-time growth rate in an unperturbed geometry. This result is actually in agrement with the switchback effect of the complexity. Moreover, according to the above calculation, we can see that the counterterm plays an important role in the slope of the complexity of formation at the large times. This is totally different from the calculation of the late-time CA complexity growth rate, where the counterterm vanishes at the late times.

\begin{figure}
\centering
\includegraphics[width=0.48\textwidth]{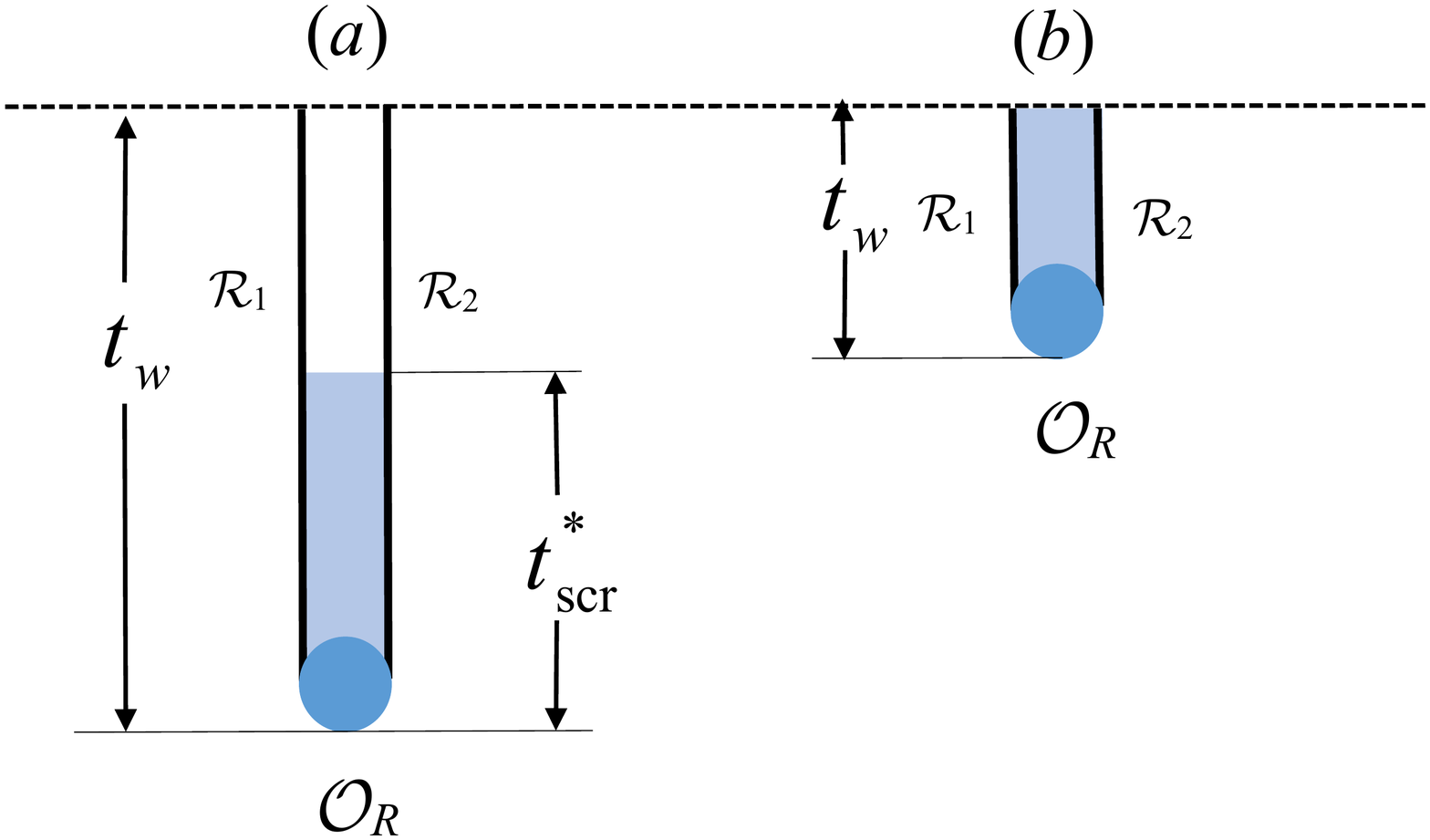}
\caption{A representation of the insertion of a  perturbed operator $\math{O}_R$ at the time $-t_w$ for the TFD state at $t_L=t_R=0$, in analogy to the construction in figure 25 of \cite{Chapman2} as well as figure 6 of \cite{D.Stanford}.}\label{swc}
\end{figure}
\section{Circuit analogy}\label{Sec4}
In this section, we would like to investigate the connection between the behaviours of our holographic results and the switchback effect of the circuit model. As discussed in Sec. \ref{Sec1}, evolving the perturbed state independently in the left and right times can be expressed as
\ba\begin{aligned}\label{SBP}
|\text{TFD}(t_L,t_R)\rangle_\text{pert}=U_R(t_R+t_w)\math{O}_RU_R(t_L-t_w)|\text{TFD}\rangle\,,\nn
\end{aligned}\nn\\\ea
where the perturbed operator $\math{O}_R$ is a localized simple operator. $U_R(t)\math{O}_R U_R(-t)=I$ with the identity operator $I$ when $t<t^*_\text{scr}$. This feature is connected to the switchback effect \cite{D.Stanford,Susskind6} and can provide a deeper explanation of our holographic results.

We denote the rate of the complexity to $\math{R}_1$ before the operator $\math{O}_R$ is inserted and $\math{R}_2$ after it\cite{Chapman2}. In the holographic context, these rates are dual to the late-time complexity growth rate of the stationary black hole. Under the limit of light shocks, we have
$
\math{R}_1\approx \math{R}_2\approx\math{R}.
$

With similar consideration as last section, here we also focus on the case $t_L=t_R=0$. Then, the complexity only depends on $t_w$. And there are two special regions which are divided by the scrambling time $t_\text{scr}^*$.

First of all, we consider the region with $t_w>t^*_\text{scr}$, where the process can be illustrated by (a) in \fig{swc}. In this case, the two time-evolution operators cancel out only during the scrambling time. Then, the complexity can be written as
\ba
C_\text{pert}\approx 2 \math{R}(t_w-t^*_\text{scr})\,.
\ea

However, for the case $t_w<t^*_\text{scr}$, as illustrated by (b) in \fig{swc}, the switchback effect produces a cancellation for the process below the dashed line. Then, the rate of the complexity vanishes.

Summarizing these results, the slope of the complexity of formation can be written as
\ba
\frac{d\D C_\text{pert}}{d t_w}\approx2 \math{R}\math{H}(t_w-t^*_\text{scr})\,.
\ea
Again, this formula is also in accord with the our holographic case as illustrated in Eqs. \eq{ec} and \eq{lc} of the last section.

\section{Conclusion}
{
In this paper, we use the CA conjecture to investigate the switchback effect of the TFD state following a quantum quench in the strongly-coupled quantum system with finite $N$ and finite coupling effects. From the viewpoint of the AdS/CFT, this quantum system is dual to a bulk gravitational theory with higher curvature corrections. Then, the investigation is equivalent to studying the switchback effect of the CA complexity in a Vaidya geometry equipped with a light shockwave.} Based on the Noether charge formalism of Iyer and Wald, a general expression can be resorting to describing the slope of the complexity of formation in the small and large $t_w$ approximations. And the large-time slope of the complexity of formation is essentially twice the late-time growth rate in an unperturbed geometry. By the circuit analogy, we showed this holographic result is essentially in agreement with the switchback effect of the quantum system. The above discussions are independent of the explicit gravitational theory as well as spacetime geometry. { This also indicates that the switchback effect is a general feature of the TFD state in the strongly-coupled system with finite $N$ and finite coupling effects. } Moreover, according to the calculation of the slope of the complexity, we can see that unlike the late-time complexity growth rate, the countertem will play an important role in the switchback effect.

\section*{acknowledge}
 This research was supported by NSFC Grants No. 11775022 and 11873044.

\end{document}